\documentclass[prl,twocolumn,preprintnumbers,superscriptaddress,
               amsmath,amssymb,aps,longbibliography,nofootinbib]{revtex4-2}

\RequirePackage[colorlinks=true,urlcolor=blue,anchorcolor=blue,
  citecolor=blue,filecolor=blue,linkcolor=blue,menucolor=blue,
  linktocpage=true,pdfproducer=medialab,pagebackref=true]{hyperref}
\hypersetup{colorlinks=true,urlcolor=blue,anchorcolor=blue,
  citecolor=blue,filecolor=blue,linkcolor=blue,menucolor=blue,
  pdfproducer=medialab}

\usepackage[utf8]{inputenc}
\usepackage{calligra,amsmath,amsfonts,bbm,mathrsfs,amssymb,mathtools}
\usepackage{slashed,cancel}
\usepackage{bm}
\usepackage{graphicx}
\usepackage{float}
\usepackage{enumerate}
\usepackage{color}
\usepackage{multirow}
\usepackage[normalem]{ulem}
\usepackage{booktabs}
\usepackage{braket}
\usepackage{MnSymbol}
\usepackage{orcidlink} 

\newif\ifinappendix
\inappendixfalse

\renewcommand{\section}[1]{%
  \ifinappendix
    \refstepcounter{section}%
    \par\bigskip
    {\centering\textbf{\thesection\space #1}\par}%
    \bigskip\noindent\ignorespaces
  \else
    \par\textbf{\textit{#1} --- }\ignorespaces
  \fi
}

\renewcommand{\appendix}{%
  \inappendixtrue
  \setcounter{section}{0}%
  \renewcommand{\thesection}{\Alph{section}}%
}
\let\vec\mathbf
\newcommand{\sumat}{\textit{Supplemental Material}}

\definecolor{rosso}{cmyk}{0,1,1,0.4}
\definecolor{rossos}{cmyk}{0,1,1,0.55}
\definecolor{rossoc}{cmyk}{0,1,1,0.2}
\definecolor{blu}{cmyk}{1,1,0,0.3}
\definecolor{blus}{cmyk}{1,1,0,0.6}
\definecolor{bluc}{cmyk}{1,1,0,0.1}
\definecolor{verde}{cmyk}{0.92,0,0.59,0.25}
\definecolor{verdec}{cmyk}{0.92,0,0.59,0.15}
\definecolor{verdes}{cmyk}{0.92,0,0.59,0.4}

\begin{document}

\title{Stochastic Tsunamis: \protect\\
Diffuse Scalar Background from Black Hole Formation}

\author{Arturo de Giorgi~\orcidlink{0000-0002-9260-5466}}
\email{arturo.de-giorgi@durham.ac.uk}
\affiliation{Institute for Particle Physics Phenomenology, Department of Physics, Durham University, Durham DH1 3LE, U.K.}

\author{Joerg Jaeckel~\orcidlink{0000-0002-6038-4785}}
\email{jjaeckel@thphys.uni-heidelberg.de}
\affiliation{Institut f\"ur Theoretische Physik, Universit\"at 
Heidelberg, Philosophenweg 16, 69120 Heidelberg, Germany}

\preprint{IPPP/26/56}

\begin{abstract}
Massive astrophysical objects can source huge static configurations of a scalar field. When such an object ends up forming a black hole, for instance, via a core-collapse supernova, the scalar field loses its source abruptly; then the static configuration becomes dynamical and propagates away in a burst, a ``scalar tsunami''. These bursts accumulate over cosmological history, forming a relic stochastic diffuse scalar background peaked in the $1-10^3$~Hz range. We propose this as a novel mechanism for the generation of such a background, compute its spectrum, and compare it with the sensitivity of future experiments. We show how this extends the experimental sensitivity to scalar masses $m_\phi\lesssim 10^{-13}$~eV, ten orders of magnitude larger than those accessible via individual transient events previously considered.
\end{abstract}

\maketitle

\renewcommand*{\thefootnote}{\arabic{footnote}}

\section{Introduction}
\label{sec:introduction}
Ultralight scalar fields are well-motivated dark matter candidates~\cite{Preskill:1982cy,Abbott:1982af,Dine:1982ah,Piazza:2010ye,Nelson:2011sf,Arias:2012az,Antypas:2022asj,Adams:2022pbo,Albertus:2026fbe,Arza:2026rsl}, and much effort has gone into studying their possible detection utilizing their dark matter background abundance. Given the absence of a detection so far, it is natural to wonder about caveats that would undermine this quest: for example, what if we are unlucky, and the local density is too low, or the field is clumped in structures far away from Earth? What if an ultra-light scalar field exists but it is a very subdominant component of dark matter? In these cases, transient signals generated by different mechanisms offer an alternative detection channel. These include the generation of transient signals from cosmic events with specific initial field configurations~\cite{Yoshino:2012kn,Eby:2016cnq,Levkov:2016rkk,Baumann:2018vus}, as well as transient signals produced via particle production~cf. e.g.~\cite{Fiorillo:2025gnd,Lecce:2025dbz,Lecce:2025vjc,Candon:2025ypl}.

Along these lines, a new mechanism was proposed in Ref.~\cite{deGiorgi:2024pjb}: scalar emission from black hole formation. Even relatively weak couplings to matter can cause large field configurations around big objects, such as stars. If the object ends up inside a black hole, e.g. in the event of a supernova or when merging with another object, any non-gauge couplings become obscured by the horizon~\cite{Israel:1967wq,Israel:1967za,Carter:1971zc,Ruffini:1971bza,Carter:2009nex,Robinson:1975bv,Carter:1979wef,Mazur:2000pn} and the field loses its source. The previously static configuration is ``released'', and part of it becomes a dynamical outward propagating ``scalar tsunami''.
In Ref.~\cite{deGiorgi:2024pjb}, the present authors identified detection opportunities considering the closest candidates to undergo such a process. The computation was then refined in Ref.~\cite{deGiorgi:2026lez}, where a full numerical analysis of the emitted spectrum was performed, including the gravitational effects of the BH remnant. Notably, this analysis also supports the naive idea that an ${\mathcal{O}}(1)$ fraction of the scalar field energy propagates outward.

As noted in Ref.~\cite{deGiorgi:2024pjb}, this detection strategy is subject to two limitations. On the one hand, transient signals of this kind only allow us to efficiently probe light scalar fields whose mass satisfies $m_\phi \lesssim 1/r$, where $r$ is the distance between Earth and the source. For larger masses, dispersion causes the signal to be too diffuse. Considering the closest sources, this translates into the largest testable mass being $m_\phi \lesssim 10^{-25}$~eV. On the other hand, we need a large amount of ``luck'', since a successful detection needs a nearby candidate to undergo such a violent phenomenon.

In this work, we take a complementary path to circumvent the just-mentioned obstacles. The question we aim to answer is: can the generation of these signals, accumulated over cosmological history, leave an imprint? If so, can future experiments use it to access a larger portion of the parameter space compared to transient events alone? It turns out that this is indeed the case. 

Relying on a persistent, stochastic signal rather than a transient one removes the limitation from dispersion.
Moreover, the accumulation of the signal over cosmological history removes the need for ``luck'', making this stochastic relic signal an irreducible contribution populating the universe today. 
Previous, examples of using such stochastic, ``diffuse'' backgrounds include axion-like and other new particles~\cite{Raffelt:2011ft,Calore:2020tjw,Calore:2021hhn,Dror:2021nyr,Eby:2024mhd}; but also neutrinos famously have a diffuse background generated by supernovae, cf., e.g.~\cite{Beacom:2010kk} which is on the verge of being detected~\cite{Super-Kamiokande:2025sxh}.
Finally, as we will discuss, the signals can be reliably computed as long as $m_\phi \lesssim 1/R$, $R$ being the radius of the source that generated it. This approach therefore allows us to probe masses up to $m_\phi \lesssim 10^{-13}$~eV.


\section{Stochastic Diffuse Background}
The log-differential abundance of a field is defined as
\begin{equation}
    \Omega(f)\equiv \frac{1}{\rho_c}\frac{d\rho}{d\text{log}f}=\frac{f}{\rho_c}\frac{d\rho}{df}\,,
\end{equation}
where $\rho_c$ is the critical energy density.
We denote by $f$ the frequency as measured today, and $f_s$ the frequency at emission.
The stochastic relic abundance generated from redshift $z_\star$ to the present day is given by cf.~\cite{Romano:2019yrj,SM},
\begin{align}
    \Omega(f)&= \frac{f}{\rho_c}\int\limits_0^{z_\star}\frac{\mathcal{R}(z)~dz}{H(z)(1+z)}\left(\frac{f(1+z)}{f_s}\right)^2\left.\frac{dE}{df_s}\right|_{f_s(f)}
\end{align}
where $H(z)$ is the Hubble rate, $dE/df_s$, is the energy spectrum emitted by the source, the redshift of the frequency is dictated by $f_s=\sqrt{f_m^2+(1+z)^2(f^2-f_m^2)}$ with $f_m\equiv m_\phi/2\pi$, and $\mathcal{R}(t)$ is the event rate per comoving volume. A pedagogical derivation can be found in the \sumat~\cite{SM}. For all numerical evaluations, we employ the central value of the cosmological parameters from the Planck~2018 results~\cite{Planck:2018vyg}.

\section{Signal Energy Spectrum}
We consider two types of initial field profiles also considered in Refs.~\cite{deGiorgi:2024pjb, deGiorgi:2026lez}: (i) a Yukawa-type profile and (ii) a ``compact'' source. 
The former one appears whenever the field couples linearly to fermion bilinears $\phi \bar{\psi}\psi$, such as nucleons or electrons, and it is characterised by a Yukawa potential profile $\phi(r)\propto e^{-m_\phi r}/r$. The latter one represents a toy model for fields with non-vanishing attractive self-interactions, which allow for more clustered static configurations. We define it as a field profile which is non-zero only within a finite size $R$; for concreteness, we choose $\phi(r)\propto (R-r)^3/R^4$.

Their energy spectrum, neglecting BH remnant effects, was computed in Ref.~\cite{deGiorgi:2024pjb}; further information can be found in the \sumat~\cite{SM}. Qualitatively, both signals are peaked around the characteristic frequency dictated by the size of the source $f_c\simeq R^{-1}$. Both signals decay rapidly at higher frequencies, whereas at lower frequencies, the Yukawa signal decays significantly more slowly than the compact one due to the longer-range potential. 

For this study, we employ the signals derived, including the gravitational effects of the produced BH, in Ref.~\cite{deGiorgi:2026lez}. Compared to the purely flat case, the signals get slightly redshifted, and the high-frequency part gets suppressed.
The massless approximations employed in Ref.~\cite{deGiorgi:2026lez} is valid for $m_{\phi}\ll R^{-1}$. For the present study, this determines the mass range we can analyse. We also expect that for higher masses the emission is likely reduced, since for masses $m_{\phi}\gtrsim 2\pi f_c$ the particles in the initial field configuration are increasingly non-relativistic and less likely to escape from the region around the BH.

We consider two case studies, where the relevant size of the source is $R=\{2,20\} R_s$, where $R_s(M)=2G_N M$ is the Schwarzschild radius of the source and $M$ its total mass. In both cases, we fix the total energy stored in the field to be a fixed fraction of the progenitor mass.\footnote{In a strict Yukawa case, this has a clear interpretation as a chosen coupling value, as one expects the total charge generating the field $\phi$ to linearly scale with the number of particles and therefore the total mass. However, parametrizing in terms of the total field energy is less dependent on the specific mechanism by which the field is sourced.} We choose the benchmark fraction of $10\%$, $E(M)=0.1 M$: different ones can be straightforwardly included by linearly rescaling the final results for the signal power. 
\begin{figure*}
    \centering
    \includegraphics[width=\linewidth]{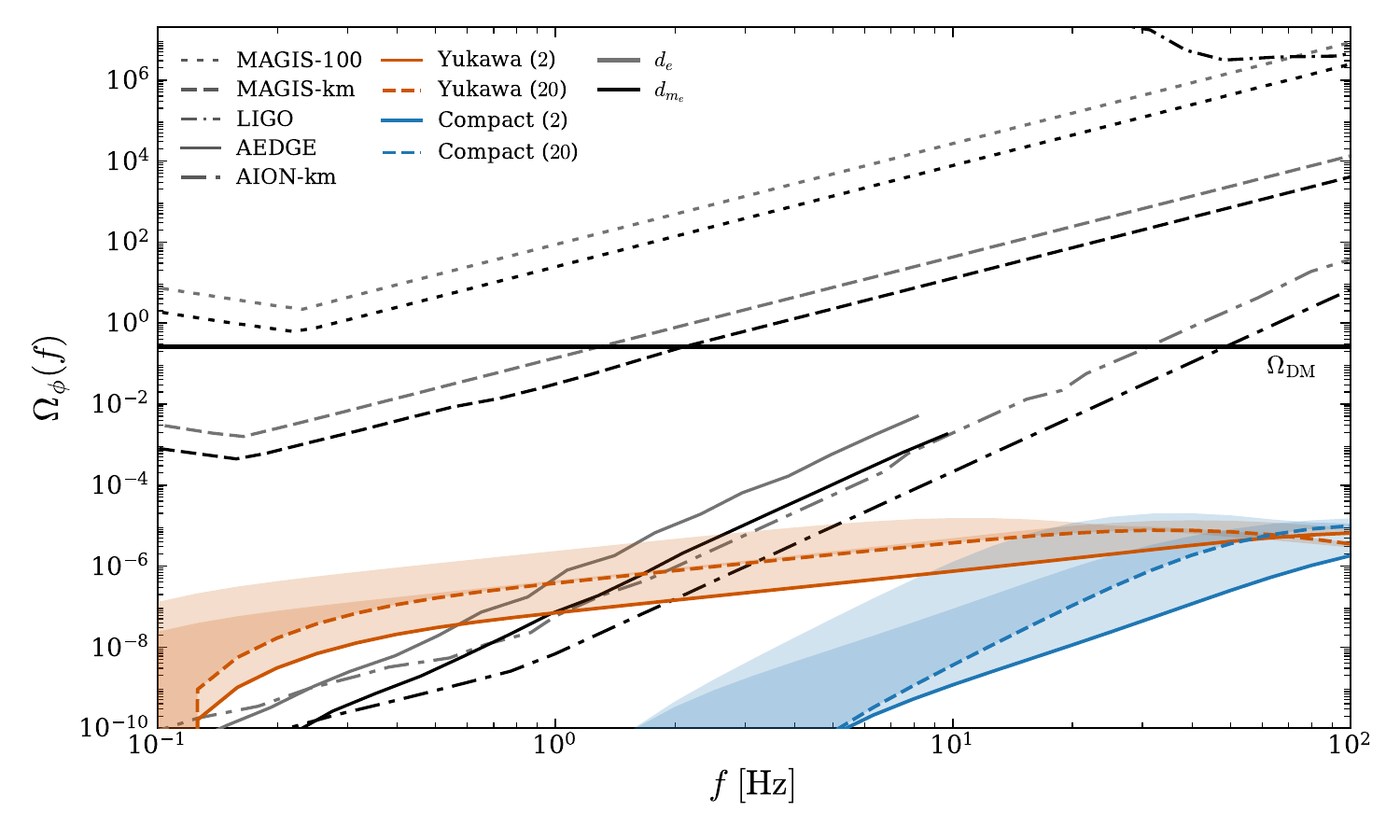}
    \caption{Signal strength and experimental sensitivity for different experiments. The couplings were set to the maximum experimentally allowed values, with the energy released per event equal to $10\%$ of the progenitor's remnant mass. The Yukawa and Compact cases are shown in orange and blue for the $R=\{2~\text{(solid)},20~\text{(dashed)}\}R_s$ cases, respectively. The spectrum is computed using a Kroupa IMF and solar metallicity. The shaded band shows the largest envelope including the metallicity uncertainty from Ref.~\cite{Ugolini:2025lzo}; the shading appears darker where the uncertainty bands of the two benchmarks overlap. For comparison, the black line highlights the DM relic density $\Omega_\text{DM}\simeq 0.26$. }
    \label{fig:experiments}
\end{figure*}
\section{Signal Sources}
\begin{table}[]
\centering
\begin{tabular}{lrr}
\toprule
\textbf{Source} & \textbf{Rate $[\text{yr}^{-1}\,\text{Gpc}^{-3}]$} & \textbf{Ref.} \\
\midrule
Core-collapse supernovae & $(0.9\text{--}1.1)\times 10^{5}$ & \cite{Taylor:2014rlo} \\
Binary neutron star mergers & $10$--$1700$ & \cite{Du:2025wto} \\
Neutron star-black hole mergers & $7.8$--$140$ & \cite{KAGRA:2021duu} \\
\bottomrule
\end{tabular}
\caption{Rates of the astrophysical sources considered as potential origins of the stochastic scalar signal.}
\label{tab:rates}
\end{table}
We identify three astrophysical processes that are potentially relevant sources of our stochastic scalar signals: core-collapse supernovae~(CCSN), binary neutron star mergers~(BNSM), and neutron star black hole mergers~(NSBH). The order-of-magnitude rates of such events are reported in Table~\ref{tab:rates}. As can be seen, CCSN events dominate, exceeding the BNSM and NSBH rates by roughly three orders of magnitude. 
In this work, we will therefore consider only the former.

The core-collapse supernova rate density, differential in the progenitor mass $M$, is directly proportional to the star formation rate~(SFR), $R_\mathrm{SF}(z)$
(see, e.g. Ref.~\cite{Calore:2021hhn}),
\begin{equation}
\label{eq:ccsnrate}
    R_\mathrm{CC}(z, M) = R_\mathrm{SF}(z)\,
    \frac{\int_{8\,M_\odot}^{125\,M_\odot} dM\,\phi(M)}
         {\int_{0.5\,M_\odot}^{125\,M_\odot} dM\, M\,\phi(M)},
\end{equation}
where $\phi(M) \propto M^{-\zeta}$ is the initial mass function~(IMF) and $M$ is the zero-age main sequence (ZAMS) mass, i.e. the mass of the star at birth. The IMF slope $\zeta$ takes values $2.35$, $2.30$, and $2.15$ for the Salpeter~\cite{Salpeter}, Kroupa~\cite{Kroupa:2000iv}, and Baldry-Glazebrook~\cite{Baldry:2003xi} IMFs, respectively, all defined for $M \gtrsim 0.5\,M_\odot$.
The denominator integrates over the full stellar mass range
$[0.5, 125]\,M_\odot$, representing the total stellar mass formed per comoving volume per unit time as traced by $R_\mathrm{SF}(z)$. The numerator counts only stars in the mass range $[8, 125]\,M_\odot$, since a minimum ZAMS mass of $\sim 8\,M_\odot$ is required for a star to undergo core collapse rather than ending its life as a white dwarf. Further details about the shape of $R_\mathrm{SF}(z)$ can be found in the \sumat~\cite{SM}.

The rate above counts core-collapse events as a function of the ZAMS mass, whereas the signal we compute requires that a BH is formed: what we need is therefore the fraction of core collapses that produce a BH, and the mass of that BH for a given progenitor. 
Importantly, not every core collapse leaves a BH behind, and the remnant mass differs substantially from the progenitor's ZAMS mass, since stars lose a metallicity-dependent fraction of their envelope to winds and further mass during the explosion. We account for this using the initial-mass to remnant-mass relation of Ref.~\cite{Ugolini:2025lzo}, obtained from hydrodynamical simulations of non-rotating progenitors, restricting the integration over $M$ to the range that produces a BH. Since this mapping depends on the poorly known metallicity of the progenitor population, we repeat our calculation for four representative metallicities and take their spread as a systematic uncertainty. Further details are given in the \sumat~\cite{SM}.

\section{Discovery Possibilities}
\label{sec:results}
In the regime of interest, the largest CCSN has a mass of $120M_\odot$, which allows us to safely constrain the $m_\phi \lesssim 10^{-13}$~eV region for the $R=2R_s$ source, and one order of magnitude less for the $R=20R_s$ case. 

A wide variety of experiments can be sensitive to a transient scalar field wave. Here, we consider scalars coupled to photons and electrons via
\begin{equation}
\mathcal{L} \supset -\phi(x,t)\cdot\sqrt{4\pi G_N} \left[d_{m_e}\,m_e\,\bar{e}e - \frac{1}{4}d_e\,F_{\mu\nu}F^{\mu\nu}\right],
\end{equation}
where $G_N$ is Newton's constant and $d_{m_e}$, $d_e$ are dimensionless couplings to the electron mass and photon kinetic term, respectively. We consider experiments looking for periodic signals originating from DM, including atom interferometers (MAGIS~\cite{MAGIS-100:2021etm}, 
AION~\cite{Badurina:2019hst}), and the gravitational wave detector LIGO~\cite{Gottel:2024cfj}. 
Cavity~\cite{Sikivie:1983ip} and lumped-circuit experiments (e.g. ADMX~\cite{Asztalos:2003px,Du:2018uak,Braine:2019fqb}, HAYSTAC~\cite{Zhong:2018rsr}, DMRadio\cite{Chaudhuri:2014dla,Silva-Feaver:2016qhh}, ABRACADABRA~\cite{Ouellet:2018beu,Ouellet:2019tlz}) probe a similar photon coupling~\cite{Dror:2021nyr}, but their sensitivity window lies above $f \gtrsim 10^5$~Hz, outside our range of interest, and are therefore not included.
Constraints are presented as curves in terms of $d_e$ or $d_{m_e}$ vs scalar mass. 

To collect all results in a single plot, we fix $d_{m_e,e}$ to the largest allowed values from non-DM experiments for a massless field. The leading experiment relevant for us is MICROSCOPE~\cite{Berge:2017ovy}, allowing the largest values of the couplings to be $d_{e,\text{max}}  = 2.3\cdot 10^{-4}$ and $d_{m_e,\text{max}}  = 10^{-3}$. 

We recast the DM constraints for our broadband signal. This is a non-trivial task, and a precise sensitivity determination requires a complete reanalysis of each experiment. Here we provide a simple estimate based on a rough scaling (see also~\cite{Dror:2021nyr}). Dark matter experiments operate under the assumption that the entire local dark matter density $\rho_{\rm DM} = 0.3~\mathrm{GeV/cm}^3$ is concentrated at a single frequency $f = m_\phi / (2\pi\hbar)$. The signal bandwidth is very small and is characterised by the quality factor $Q_\text{DM}\equiv f/\Delta f_\text{DM} \simeq 10^{6}$, reflecting the narrow velocity dispersion of the halo.
In contrast, our stochastic background is broad in frequency. To compare the signal to that of dark matter, we estimate the energy in the diffuse background in the same frequency interval where the DM is searched for, $[f,f(1+Q_{\text{DM}})]$,
\begin{equation}
    \Delta\rho_\phi(f)= \rho_c \int^{\log(1+1/Q_{DM})}_{\log(1)}d\log(f)\,\Omega_\phi(f)
    \simeq \rho_c \frac{\Omega_\phi(f)}{Q_{\text{DM}}}\,,
\end{equation}
where $Q_{\text{DM}}\sim 1/v^{2}_{\text{DM}}\sim 10^6$ quantifies the width of the DM signal.
Comparing the signal strength in the two cases, we naively obtain
\begin{equation}
    \frac{\Delta\rho_\phi\left(f=\frac{m_\text{DM}}{2\pi}\right)}{\rho_\text{DM}}\simeq \frac{\Omega_\phi\left(\frac{m_\text{DM}}{2\pi}\right)}{\Omega_\text{DM}}\times\frac{1}{Q_\text{DM}}\,.
\end{equation}
As noted in Ref.~\cite{Dror:2021nyr}, the above rescaling is overly pessimistic. Since the broad signal occupies $\sim Q_\text{DM}$ independent frequency bins simultaneously, combining them yields a statistical enhancement of $\sqrt{Q_\text{DM}}$ in the signal-to-noise ratio, so that the naive $1/Q_\text{DM}$ suppression is replaced by only a $\sqrt{1/Q_\text{DM}}$ penalty in sensitivity. Therefore, up to an $\mathcal{O}(1)$ factor, the sensitivity features only a square-root dependence on the quality factor
\begin{equation}
\label{eq:rescaling}
    \Omega_\text{sens}(f)\simeq\Omega_\text{DM}\times\sqrt{Q_\text{DM}}\times\left(\frac{d_\text{exp}(f)}{d_\text{max}}\right)^2\,,
\end{equation}
where $d_\text{max}$ is the DM-independent reference coupling in question, and $d_\text{exp}(f)$ is the frequency-dependent bound from the DM experiments.
Of course, this combination of independent frequency bins can only be used if the experiment is simultaneously sensitive over the combined frequency range. This is indeed the case for the experiments considered below. 

A further subtlety, not captured by the above rescaling, concerns relativistic effects: if the signal wavelength $\lambda= 1/f$ becomes comparable to the size $R$ of the detector, the field can vary appreciably across the apparatus, partially averaging out the signal (see Ref.~\cite{Dror:2021nyr} for the discussion of such an effect for resonant cavity haloscopes). For all the experiments considered here, we have $\lambda\gg R$ over essentially the full frequency range entering our sensitivity projections. The only marginal exception is AEDGE at its highest frequencies. However, in that regime, the experiment has no sensitivity regardless. The shortest baseline considered, MAGIS-100 ($R\simeq100$~m), only requires $f\lesssim 3\times10^{6}$~Hz, while the longest, AEDGE ($R\simeq1000$~km), is the most restrictive, requiring $f\lesssim 3\times10^{2}$~Hz. To remain conservative, we restrict the highest frequency we consider to be $10^2$~Hz, ensuring $\lambda\gg R$ for all experiments. 

The sensitivity curves of the different experiments are shown in Fig.~\ref{fig:experiments}.  The sensitivity curves for the $d_e$ and $d_{m_e}$ couplings are shown for MAGIS-100, MAGIS-km, AEDGE, AION-km and LIGO. We show results for two source models: a Yukawa profile (orange) and a compact object profile (blue), each for two choices of the source radius $R = \{2, 20\}\,R_s$, where $R_s$ is the Schwarzschild radius. As mentioned, we fix the energy stored in the field to be $10\%$ of the progenitor's mass. 
The fiducial spectrum is computed assuming a Kroupa initial mass function and a solar-metallicity progenitor population; the shaded bands reflect the systematic uncertainty from the initial-mass to remnant-mass mapping, obtained by varying the progenitor metallicity. As can be seen, future detectors with improved sensitivity, in particular MAGIS-km and AEDGE, have sensitivity to the Yukawa case at low frequencies.


\section{Conclusions and Outlook}
In this work, we investigated whether ultralight scalar fields, sourced by massive astrophysical objects that collapse into black holes~(BH), can leave a stochastic imprint accumulated over cosmological history, complementing the search for transient events. The idea is that each such collapse emits a burst of scalar radiation, a ``scalar tsunami'', and the superposition of many such bursts throughout cosmic history builds up a stochastic background of scalar radiation permeating the universe today.

Our main result is shown in Fig.~\ref{fig:experiments}, where we compare the predicted stochastic signal to the projected sensitivities of current and future experiments. We find that this relic background falls in the frequency domain that can be probed by atom interferometers such as MAGIS-100, MAGIS-km, AEDGE, and AION-km, as well as by gravitational wave detectors such as LIGO. The signal is approximately mass-independent for masses up to $m_\phi \lesssim 10^{-13}$~eV. This extends the reach of transient searches to a region of parameter space that would otherwise require a rare, extremely nearby source.

Extending this analysis to larger scalar masses is a natural next step, though it requires a more detailed modelling of the BH source, as the emission spectrum becomes more sensitive to the structure of the collapsing object and the interplay between the scalar Compton wavelength and the Schwarzschild radius. This is likely to reduce the fraction of scalars escaping from the BH. However, heavier scalars also offer a compelling opportunity: once the emitted radiation redshifts into the non-relativistic regime, the energy density accumulates near the rest-mass frequency $f \simeq m_\phi/(2\pi\hbar)$, concentrating the signal into a narrow spectral feature that dark matter direct detection experiments are optimised for, potentially enhancing the prospects for discovery.
\section{Acknowledgments}
AdG thanks G.~Lucente, F.~Silvetti, and O.~Straniero for useful discussions and comments. 
JJ would like to thank X.~Ma and V.~Takhistov for fun collaboration on related topics. A.d.G. acknowledges support from the COST Action COSMIC WISPers CA21106, supported by COST (European Cooperation in Science and Technology).

\bibliographystyle{BiblioStyle}
\bibliography{Bibliography}
\newpage
\appendix

\begin{center}
  {\large \textbf{Supplemental Material}
  }
\end{center}

\renewcommand{\thepage}{S\arabic{page}}
\setcounter{page}{1}

\renewcommand{\theequation}{S\arabic{equation}}
\setcounter{equation}{0}

In this \sumat, we provide further details about the energy spectra and the parametrisation of the core collapse supernova event rate. 

\section{Flat Space Energy Spectra}
\label{sec:energy-spectra}
For the computation of the diffuse background, it is crucial to estimate the energy spectrum emitted from the source. Below, we report the energy spectrum as a function of the momentum for both the Yukawa and Compact cases in flat space-time, i.e., neglecting the BH remnant effect.
From that, the spectrum in frequency can be obtained straightforwardly by employing $\omega_k=\sqrt{m_\phi^2+k^2}=2\pi f$ and
\begin{equation}
    \label{eq:k-f-relation}\frac{dE}{df}=2\pi \frac{dE}{d\omega_k}=2\pi \frac{\omega_k}{k} \times\frac{dE}{dk}\,.
\end{equation}
The expressions are normalised such that the total energy $E$ is obtained upon
\begin{equation}
    E=\int\limits_0^\infty dk~\frac{dE}{dk}\,.
\end{equation}

\paragraph{Yukawa Source.}
The value of $\phi(r,t)$ in flat space, as well as its emitted energy distribution in momentum $dE/dk$ were derived in Ref.~\cite{deGiorgi:2024pjb}. 
For a Yukawa-type massive field, it reads
\begin{align}
    &\frac{dE}{dk}=\frac{9g^2}{4\pi^2 R^4}\frac{\mathcal{F}(k R)^2}{\omega_k^2 k^2}\,, \\
    &E=\frac{3g^2}{16\pi x^5 R}\left[(3-3x^2+2x^3)-3e^{-2x}(1+x)^2\right]\,.
\end{align}
where $R$ is the radius of the progenitor, $x\equiv m_\phi R$, and
\begin{equation}
    \mathcal{F}(x)\equiv \frac{\sin(x)}{x}-\cos(x)\,.
\end{equation}
In terms of frequency, $\omega_k=2\pi f$,
\begin{align}
    \frac{dE}{df}&=\left.\frac{9g^2}{2\pi R^4}\frac{\mathcal{F}(k R)^2}{\omega_k k^3}\right|_{k=k(f)}\,.
\end{align}

\paragraph{Compact Source}
For the compact source, we take it to be non-zero in a finite domain
\begin{align}
    &\phi_\text{CS}(x)=\begin{cases}
        \phi_\text{CS}(r) & r\leq R\,,\\
        0 &\text{otherwise}\,,
    \end{cases}
\end{align}
and to be $C^2$-differentiable to simplify the discussion, avoiding discontinuities in the Klein-Gordon equation with the boundary condition
\begin{equation}
    \phi(R)=\phi'(R)=\phi''(R)=0\,.
\end{equation}
The simplest function that can be taken is a third-order polynomial of the form
\begin{equation}
\label{eq:CS-benchmark}
    \phi_\text{CS}(r) = \sqrt{\frac{630|E|R}{108\pi+5\pi m_\phi^2 R^2}} \frac{(R-r)^3}{R^4}\,,
\end{equation}
where $E$ is the energy stored in the field. The differential energy spectrum is given by
\begin{widetext}
\begin{equation}
    \frac{dE}{dk}=\left(\frac{90720~|E|R}{108\pi+5\pi m_\phi^2 R^2}\right)~\frac{\omega_k^2}{R^8k^{10}}\left[(kR)^2+kR\sin(kR)+4\cos(kR)-4\right]^2\,.
\end{equation}
\end{widetext}

\section{Relic Stochastic Abundance}
\label{sec:abundance}
We derive here the main formulas to compute the energy spectrum of the signal (cf.~\cite{Romano:2019yrj} for some useful literature).~\footnote{For a complementary discussion, including some generalizations, see~\cite{Ma}.} We consider first a purely massless field. We then generalise it to the massive case. 

First, let us fix some notation and relations that will be useful. Denoting by $a_0$ the current scale factor, the redshift $z$ is given by
\begin{align}
    &1+z=\frac{a_0}{a}\,, && da=-\frac{a_0}{(1+z)^2}dz\,, && dt=-\frac{dz}{H(z)(1+z)}\,.
\end{align}

For a scalar field in an expanding universe, the field evolves as
\begin{equation}
    \ddot\phi +3H\dot\phi -\frac{1}{a^2}\nabla^2\phi+m_\phi^2\phi=0\,.
\end{equation}
By employing the Fourier decomposition
\begin{align}
    &\phi(\Vec{x},t)=\int\limits\frac{d^3k}{(2\pi)^3}\left(a_{\vec{k}}u(t)e^{i\vec{k}\cdot\vec{x}}+\text{h.c.}\right)\,, 
    \\\nonumber
    &\qquad\qquad u(t)=a^{-3/2}\chi(t)\,,
\end{align}
one finds
\begin{equation}
    \ddot\chi+\left[m_\phi^2+\left(\frac{k}{a}\right)^2-\frac{3}{2}\dot H-\frac{9}{4}H^2\right]\chi=0\,,
\end{equation}
yielding the effective dispersion relation
\begin{equation}
    \omega(k)^2=m_\phi^2+\left(\frac{k}{a}\right)^2-\frac{3}{2}\dot H-\frac{9}{4}H^2\,.
\end{equation}
For sub-horizon modes, $m_\phi^2+\left(\frac{k}{a}\right)^2\gg H^2,\dot H$, and thus $\omega(k)^2\approx m_\phi^2+\left(\frac{k}{a}\right)^2$, and the expansion just redshifts the momentum. We will work in this regime. The energy at redshift $z$ is then
\begin{align}
    \label{energy-redshift}&\omega(z)=\sqrt{m_\phi^2+(1+z)^2(\omega_0^2-m_\phi^2)}\,, 
    \\\nonumber
    &\quad\omega_0=\sqrt{m_\phi^2+\frac{\omega^2-m_\phi^2}{(1+z)^2}}\,.
\end{align}

In all cases, we are interested in computing the spectral energy density today
\begin{equation}
    \Omega_0(f_0)\equiv \frac{1}{\rho_c}\frac{d\rho_0}{d\text{log}f_0}=\frac{f_0}{\rho_c}\frac{d\rho_0}{df_0}\,,
\end{equation}
where $f_0$ is the frequency today and $f_s$ is the one emitted at the source.
For a fixed redshift, the contribution to today's abundance is given by
\begin{equation}
    \left.\frac{d\rho_0}{df_0}\right|_z= dn_0(z) \frac{dE_0}{df_0}=dn_c(z) \frac{dE_0}{df_0}\,,
\end{equation}
where $dn_0\equiv dn_c$ is the comoving number density, which is conventionally fixed to the current volume.
By making use of the conservation of the number of particles
\begin{equation}
    dN=\frac{1}{\omega_0}\frac{dE_0}{df_0}df_0=\frac{1}{\omega_s}\frac{dE}{df_s}df_s\,,
\end{equation}
we can relate the current and emitted spectra
\begin{equation}
    \frac{dE_0}{df_0}=\frac{f_0}{f_s}\left(\frac{df_s}{df_0}\right)\frac{dE}{df_s}\,.
\end{equation}
Denoting by $f_m\equiv m/2\pi$, and employing Eq.~\eqref{energy-redshift}, the Jacobian is given by
\begin{align}
   &f_s=\sqrt{f_m^2+(1+z)^2(f_0^2-f_m^2)}\,, &\frac{df_s}{df_0}=(1+z)^2\frac{f_0}{f_s}\,.
\end{align}
The supernova rate is
\begin{equation}
    \mathcal{R}(t)\equiv \frac{\text{Number of supernova}}{dt dV_c}\equiv \frac{dn_c}{dt}\,,
\end{equation}
where $V_c$ is the comoving volume. Therefore,
\begin{equation}
    dn_c= \mathcal{R}(t) dt=-\mathcal{R}(z)\frac{dz}{H(z)(1+z)}\,.
\end{equation}
All in all, we find
\begin{align}
    \Omega_0(f_0)&= \frac{f_0}{\rho_c}\int\limits_0^{z_\star}dz~\frac{\mathcal{R}(z)}{H(z)(1+z)}\left(\frac{f_0 (1+z)}{f_s}\right)^2\left.\frac{dE}{df_s}\right|_{f_s(f_0)}\,.
\end{align}
In the main text, we use this result, renaming $f_{0}\to f$ for simplicity.

We employ the following central values of the cosmological parameters from the Planck~2018 results~\cite{Planck:2018vyg}
\begin{align}
    &H_0=67.66\pm 0.42~\left[\frac{\text{Km}}{\text{s}~\text{Mpc}}\right] \,,\\ &\rho_c=\frac{3H_0^2}{8\pi G}\approx 8.597\times 10^{-27}~\left[\frac{\text{Kg}}{\text{m}^3}\right]\,.
\end{align}

\section{Sources of Uncertainty}

In this section, we briefly discuss the uncertainties associated with the modelling of the supernova rates and properties.

\subsection{Event Rates}
As mentioned in the main text, we initially consider three astrophysical processes as potential sources of stochastic scalar signals: core-collapse supernovae (CCSN), binary neutron star mergers and neutron star black hole mergers. However, their rates differ by roughly three and four orders of magnitude, respectively, $\mathcal{R}_\text{CCSN} \sim \mathcal{O}(10^5)\,\text{yr}^{-1}\,\text{Gpc}^{-3}$~\cite{Rosca-Mead:2023tdc}, $\mathcal{R}_\text{BNSM} \sim \mathcal{O}(10^2)\,\text{yr}^{-1}\,\text{Gpc}^{-3}$~\cite{Du:2025wto} and $\mathcal{R}_\text{NSBH} \sim \mathcal{O}(10^1)\,\text{yr}^{-1}\,\text{Gpc}^{-3}$\cite{KAGRA:2021duu}. Therefore, CCSN give the dominant contribution to the signal. We will therefore only discuss CCSN supernovae in the following.

As in Eq.~\eqref{eq:ccsnrate} of the main text, the CCSN rate per progenitor mass $M$ interval is given by
(see Ref.~\cite{Calore:2021hhn} whom we follow),
\begin{equation}
    R_\mathrm{CC}(z, M) = R_\mathrm{SF}(z)\,
    \frac{\int_{8\,M_\odot}^{125\,M_\odot} dM\,\phi(M)}
         {\int_{0.5\,M_\odot}^{125\,M_\odot} dM\, M\,\phi(M)},
    \label{eq:RCC}
\end{equation}
where, $R_\mathrm{SF}(z)$ is the star formation rate and $\phi(M)$ is the initial mass function.
The star formation rate~(SFR) is described by the parametric fit of~\cite{Yuksel:2008cu},
\begin{equation}
    R_\mathrm{SF}(z) = R_\mathrm{SF}(0)
    \left[
        (1+z)^{\alpha\eta} 
        + \left(\frac{1+z}{B}\right)^{\beta\eta}
        + \left(\frac{1+z}{D}\right)^{\gamma\eta}
    \right]^{1/\eta},
    \label{eq:SFR}
\end{equation}
where $R_\mathrm{SF}(0)$ is the local normalization, $B$ and $D$ encode 
the redshift breaks, $\eta \simeq -10$ controls the smoothness of the 
transitions, and $\alpha$, $\beta$, $\gamma$ are the logarithmic slopes 
at low, intermediate, and high redshift, respectively.
The redshift break 
coefficients $B$ and $D$ in Eq.~\eqref{eq:SFR} are not free 
parameters but are derived from the transition redshifts 
$z_1$ and $z_2$ and the spectral indices
\begin{equation}
    B = (1+z_1)^{1-\alpha/\beta}, \qquad 
    D = (1+z_1)^{(\beta-\alpha)/\gamma}(1+z_2)^{1-\beta/\gamma}.
    \label{eq:BD}
\end{equation}
The numerical values of all SFR parameters are reported 
in Table~\ref{tab:SFR}, taken from~\cite{Horiuchi:2008jz}.
\begin{table}[h]
    \centering
    \begin{tabular}{lcccccc}
        \hline\hline
        Model & $R_\mathrm{SF}(0)$ & $\alpha$ & $\beta$ & $\gamma$ & $z_1$ & $z_2$ \\
        \hline
        Upper    & 0.0213 & 3.6 & $-0.1$ & $-2.5$ & 1 & 4 \\
        Fiducial & 0.0178 & 3.4 & $-0.3$ & $-3.5$ & 1 & 4 \\
        Lower    & 0.0142 & 3.2 & $-0.5$ & $-4.5$ & 1 & 4 \\
        \hline\hline
    \end{tabular}
    \caption{Parameters of the SFR model of Eq.~\eqref{eq:SFR}, 
    taken from~\cite{Horiuchi:2008jz}. The normalization 
    $R_\mathrm{SF}(0)$ is in units of $M_\odot\,\mathrm{yr}^{-1}\,\mathrm{Mpc}^{-3}$.}
    \label{tab:SFR}
\end{table}

\subsection{Remnant Black Hole Masses from CCSN}
The rate $R_\mathrm{CC}(z,M)$ in Eq.~\eqref{eq:RCC} counts core-collapse events as a function of the progenitor's zero-age main-sequence (ZAMS) mass $M$. The ZAMS mass is the mass of the star when it was ``born'', way before the CCSN event. The signal, however, is sourced by the black hole~(BH) that the collapse leaves behind, so what is needed is the fraction of collapses that produce a BH, and the corresponding BH mass, as a function of $M$. Neither follows trivially from the ZAMS mass~\cite{Sukhbold:2015wba, Maraston_2025,Ugolini:2025lzo}. First, only progenitors above a threshold of $\sim 16$--$20\,M_\odot$ leave a BH, either through fallback of matter onto the proto-compact object after a successful explosion or through the direct collapse of the star when the explosion fails; lighter progenitors leave a neutron star instead. Second, the star arrives at collapse considerably lighter than at birth, having shed mass through metal-line-driven winds throughout its life, and part of the residual envelope is further ejected during the explosion. Because wind efficiency scales with the metal content of the stellar plasma, both effects depend strongly on metallicity: at solar metallicity a $120\,M_\odot$ progenitor is stripped down to a BH of only $\sim 28\,M_\odot$, whereas at low iron abundance, $[\mathrm{Fe/H}] \lesssim -2$ (with $[\mathrm{Fe/H}] \equiv \log_{10}(N_\mathrm{Fe}/N_\mathrm{H})_\star - \log_{10}(N_\mathrm{Fe}/N_\mathrm{H})_\odot$ the logarithmic iron abundance relative to solar, so that $[\mathrm{Fe/H}]=0$ denotes solar metallicity), winds are quenched and stars of $\sim 80\,M_\odot$ collapse almost entirely into BHs of comparable mass. The mapping between progenitor and remnant has been the subject of intense study (see e.g. Refs.~\cite{Sukhbold:2015wba, Maraston_2025,Ugolini:2025lzo}), and remains model dependent, the dominant uncertainty being the explodability of the progenitor, set by the structure of its pre-supernova core.

\begin{figure}
    \centering
    \includegraphics[width=0.90\columnwidth]{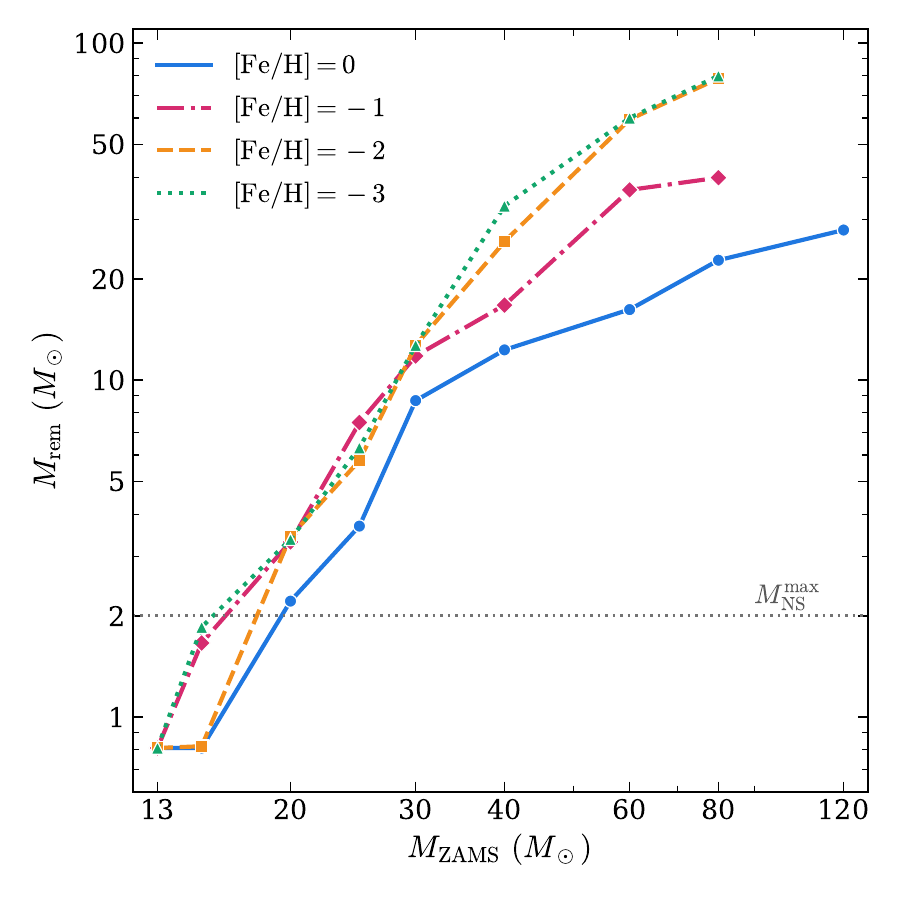}
    \caption{Remnant BH mass as a function of the progenitor ZAMS mass, from the hydrodynamical simulations of non-rotating progenitors from Table C.2 of Ref.~\cite{Ugolini:2025lzo}, for four values of the initial metallicity. The horizontal dotted line marks the maximum neutron star mass, $M_\mathrm{NS}^\mathrm{max} = 2\,M_\odot$; smaller remnants do not end in a BH and do not contribute to our signal. Curves at subsolar metallicity terminate at $80\,M_\odot$, above which progenitors become unstable.}
    \label{fig:mrem}
\end{figure}

Here we adopt the initial-mass to remnant-mass relation of Ref.~\cite{Ugolini:2025lzo}, obtained from one-dimensional hydrodynamical simulations of non-rotating progenitors, which provides the remnant mass on a grid of ZAMS masses for four metallicities, $[\mathrm{Fe/H}] = 0, -1, -2, -3$. The relation is shown in Fig.~\ref{fig:mrem}: the remnant mass grows steeply with the ZAMS mass up to $\sim 30\,M_\odot$, roughly independently of metallicity, while at larger masses the curves fan out, with metal-poor progenitors retaining most of their mass and metal-rich ones being heavily stripped by winds.

The figure also shows the two limits of validity of the mapping: the crossing of the maximum neutron star mass at low ZAMS masses, below which no BH forms, and the onset of pair instability at high masses (roughly where the lines terminate). In practice, we replace the ZAMS mass with the remnant mass $M_\mathrm{rem}(M)$ as the BH mass entering the emission spectrum, and restrict the integration over $M$ in Eq.~\eqref{eq:RCC} to the BH-forming range. Since the cosmic metallicity evolves with redshift, the appropriate relation would in principle vary along the $R_\mathrm{SF}(z)$ integration. Rather than convolving with an uncertain metallicity distribution, we calculate results separately for each fixed metallicity, and interpret their spread as the systematic uncertainty associated with the progenitor-to-remnant mapping.


\end{document}